\documentstyle[epsfig]{mn}

\newif\ifAMStwofonts

\ifoldfss
  \ifCUPmtlplainloaded \else
    \NewTextAlphabet{textbfit} {cmbxti10} {}
    \NewTextAlphabet{textbfss} {cmssbx10} {}
    \NewMathAlphabet{mathbfit} {cmbxti10} {} % for math mode
    \NewMathAlphabet{mathbfss} {cmssbx10} {} %  "   "    "
  \fi
  \ifAMStwofonts
    \ifCUPmtlplainloaded \else
      \NewSymbolFont{upmath} {eurm10}
      \NewSymbolFont{AMSa} {msam10}
      \NewMathSymbol{\upi}     {0}{upmath}{19}
      \NewMathSymbol{\umu}     {0}{upmath}{16}
      \NewMathSymbol{\upartial}{0}{upmath}{40}
      \NewMathSymbol{\leqslant}{3}{AMSa}{36}
      \NewMathSymbol{\geqslant}{3}{AMSa}{3E}

    \fi
  \fi
\fi % End of OFSS

\ifnfssone
  \newmathalphabet{\mathit}
  \addtoversion{normal}{\mathit}{cmr}{m}{it}
  \addtoversion{bold}{\mathit}{cmr}{bx}{it}
  \newmathalphabet{\mathbfit} % math mode version of \textbfit{..}
  \addtoversion{normal}{\mathbfit}{cmr}{bx}{it}
  \addtoversion{bold}{\mathbfit}{cmr}{bx}{it}
  \newmathalphabet{\mathbfss} % math mode version of \textbfss{..}
  \addtoversion{normal}{\mathbfss}{cmss}{bx}{n}
  \addtoversion{bold}{\mathbfss}{cmss}{bx}{n}
  \ifAMStwofonts
    \ifCUPmtlplainloaded \else
      \UseAMStwoboldmath
      \makeatletter
      \new@mathgroup\upmath@group
      \define@mathgroup\mv@normal\upmath@group{eur}{m}{n}
      \define@mathgroup\mv@bold\upmath@group{eur}{b}{n}
      \edef\UPM{\hexnumber\upmath@group}
      \new@mathgroup\amsa@group
      \define@mathgroup\mv@normal\amsa@group{msa}{m}{n}
      \define@mathgroup\mv@bold\amsa@group{msa}{m}{n}
      \edef\AMSa{\hexnumber\amsa@group}
      \makeatother
      \mathchardef\upi="0\UPM19
      \mathchardef\umu="0\UPM16
      \mathchardef\upartial="0\UPM40
      \mathchardef\leqslant="3\AMSa36
      \mathchardef\geqslant="3\AMSa3E
    \fi
  \fi
\fi % End of NFSS release 1

\ifnfsstwo
  \DeclareMathAlphabet{\mathbfit}{OT1}{cmr}{bx}{it}
  \SetMathAlphabet\mathbfit{bold}{OT1}{cmr}{bx}{it}
  \DeclareMathAlphabet{\mathbfss}{OT1}{cmss}{bx}{n}
  \SetMathAlphabet\mathbfss{bold}{OT1}{cmss}{bx}{n}
  \ifAMStwofonts
    \ifCUPmtlplainloaded \else
      \DeclareSymbolFont{UPM}{U}{eur}{m}{n}
      \SetSymbolFont{UPM}{bold}{U}{eur}{b}{n}
      \DeclareSymbolFont{AMSa}{U}{msa}{m}{n}
      \DeclareMathSymbol{\upi}{0}{UPM}{"19}
      \DeclareMathSymbol{\umu}{0}{UPM}{"16}
      \DeclareMathSymbol{\upartial}{0}{UPM}{"40}
      \DeclareMathSymbol{\leqslant}{3}{AMSa}{"36}
      \DeclareMathSymbol{\geqslant}{3}{AMSa}{"3E}
    \fi
  \fi
\fi % End of NFSS release 2

\ifCUPmtlplainloaded \else
  \ifAMStwofonts \else % If no AMS fonts
    \def\upi{\pi}
    \def\umu{\mu}
    \def\upartial{\partial}
  \fi
\fi

\def\gsim{\lower.73ex\hbox{$\sim$}\llap{\raise.4ex\hbox{$>$}}$\,$}
\def\lsim{\lower.73ex\hbox{$\sim$}\llap{\raise.4ex\hbox{$<$}}$\,$}

\def\%{~per~cent}

\title{Constraints on Cardassian Expansion}
\author[W.J. Frith]
{W.J.~Frith \\
Department of Physics, University of Durham, Science Laboratories, South Road, Durham, DH1 3LE, United Kingdom\\}

\pagerange{\pageref{firstpage}--\pageref{lastpage}}
\pubyear{2003}

\begin{document}

\maketitle

\label{firstpage}

\begin{abstract}
High redshift supernovae and Cosmic Microwave Background data are used to constrain the 
Cardassian expansion model (Freese \& Lewis, 2002), a cosmology in which a modification to the Friedmann equation gives 
rise to a flat, matter-dominated Universe which is undergoing a phase of accelerated expansion. In particular, the 
precision of the positions of the Doppler peaks in the CMB angular power spectrum provided by the Wilkinson Microwave 
Anisotropy Probe (WMAP) tightly constrains the cosmology. The available parameter space is further constrained by 
various high redshift supernova datasets taken from Tonry et al. (2003), a sample of 230 supernovae collated 
from the literature, in which fits to the distance and extinction have been recomputed where possible and a consistent
zero-point has been applied. In addition, the Cardassian model can also be loosely constrained by inferred upper limits on the
epoch at which the Cardassian term in the modified Friedmann equation begins to dominate the expansion ($z_{eq}$). 
Using these methods, a Cardassian cosmology is
constrained at the 2$\sigma$ level to 0.19$<\Omega_m$\lsim0.26, 0.01$<n<$0.24 for the Cardassian expansion parameter, $n$, 
and 0.42$<z_{eq}<$0.89, in contradiction to the previous constraints of Sen \& Sen (2003a). There is also a large discrepancy 
between the 1$\sigma$ confidence regions defined by the CMB and tightest Supernova constraints, with the CMB data favouring 
a low-$\Omega_m$, high-$n$ Cardassian cosmology and $z_{eq}>$1, as opposed to the Supernova data which supports a 
high-$\Omega_m$, low-$n$ cosmology. 
\end{abstract}

\section{Introduction}

Over the last ten years, the $m$-$z$ relation below $z\approx$1 has been constrained through observations of distant type Ia
supernovae, and indicates that the Universe is currently in a phase of accelerated expansion (e.g. Perlmutter et al. 1997; 
Riess et al. 1998). This is supported by recent measurements of the Cosmic Microwave Background angular power spectrum and
the galaxy power spectrum. The standard model provides an interpretation of these results by proposing that the Universe
consists of only $\sim$4\% baryonic matter with the remainder in the form of Cold Dark Matter and a dark energy component 
which drives the expansion. This cosmology provides a good fit to the Wilkinson Microwave Anisotropy Probe (WMAP) CMB power
spectrum and high redshift supernova data, and is consistent with local large-scale structure and cluster baryon
fraction measurements of the matter density. This cosmology is slightly problematic however as it invokes the use of two
unobserved forms of energy. While the evidence for Cold Dark Matter is considerable, evidence for the 
existence of a dark energy component remains indirect.

An interesting alternative to the standard model of cosmology recently proposed by Freese \& Lewis (2002) invokes an
additional term in the Friedmann equation, proportional to $\rho^n$, which results in a Universe which is
consistent with the recent evidence for an increasing expansion rate, but is both flat and matter dominated. The need
for a dark energy component is therefore removed, and the expansion is driven solely by the new $\rho^n$ term in the
modified Friedmann equation. The theoretical motivation for this Cardassian model of cosmology (see Freese \& Lewis 
2002) is fairly speculative however, and the implications for Einstein's equations of General Relativity are currently 
undetermined.

In order to constrain the Cardassian model, it is critical that the observational data used do not assume a prior
cosmology. The simplest constraints can therefore be imposed by the CMB angular power spectrum, high redshift supernova
data and local large-scale structure, since in a Cardassian cosmology these observables depend on $\Omega_m$,
the Cardassian parameter $n$, and the redshift at which the Cardassian term in the modified Friedmann equation begins to
dominate, $z_{eq}$.

Previously, Sen \& Sen (2003a) used predictions for the locations of the first and third Doppler peaks in
the CMB power spectrum for a Cardassian cosmology, and Archeops and Boomerang data to constrain possible values
for $\Omega_m$ and $n$. Supernova data from the Calan-Tololo project and the Supernova Cosmology Project (Perlmutter et
al. 1997) were also used to constrain these parameters, since the modification of the luminosity distance in a Cardassian
cosmology alters the predicted apparent magnitude for supernovae. These two constraints restricted the parameters
to 0.31\lsim$n$\lsim0.44 and 0.13\lsim$\Omega_m$\lsim0.23. Zhu \& Fujimoto (2003) also used these supernova data to
constrain the Cardassian model, and found best fit parameters of $n$=-1.33 and $z_{eq}$=0.43 assuming $\Omega_m$=0.3.
However, it is not clear how the fit is affected for different values of $\Omega_m$. More recently Sen \& Sen (2003b) 
used WMAP and Boomerang measurements of the locations for the first, second and third Doppler peaks to further constrain 
the parameter space. 

In this paper, the parameter space is constrained using the WMAP and Boomerang measurements for the locations of the 
first, second and third Doppler peaks in the CMB angular power spectrum used in Sen \& Sen (2003b), and various 
high redshift supernova datasets taken from Tonry et al. (2003). An orthogonal constraint is also applied using an 
assumed upper limit of $z_{eq}$=1; this value is used in order to be consistent with local large-scale structure 
(Freese \& Lewis, 2002). In Section 2, the Cardassian model is outlined. The supernova analysis method and the resulting 
constraints for various samples are presented in Section 3, along with the CMB constraints and the additional constraint 
imposed by the inferred upper limit on $z_{eq}$. The discussion and conclusions follow in Section 4.

\begin{figure}
\begin{center}
\centerline{\epsfxsize = 3in
\epsfbox{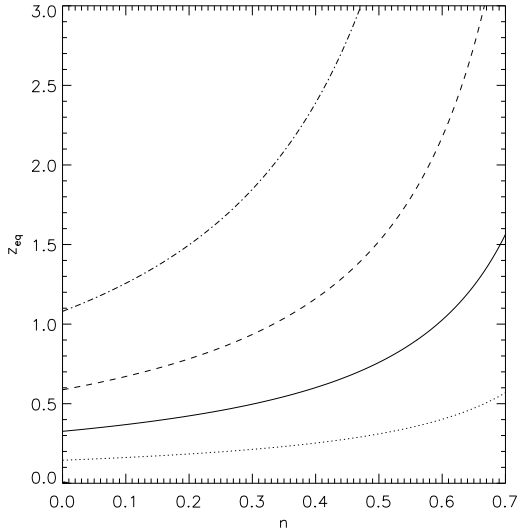}}  
\caption{Contours of $\Omega_m$ with the Cardassian parameters $n$ and $z_{eq}$. $\Omega_m$=0.1 (dot-dashed), $\Omega_m$=0.2
(dashed), $\Omega_m$=0.3 (solid) and $\Omega_m$=0.4 (dotted) tracks are shown.}
\label{fig:f_fact}
\end{center}
\end{figure}

\begin{figure}
\begin{center}
\centerline{\epsfxsize = 3in
\epsfbox{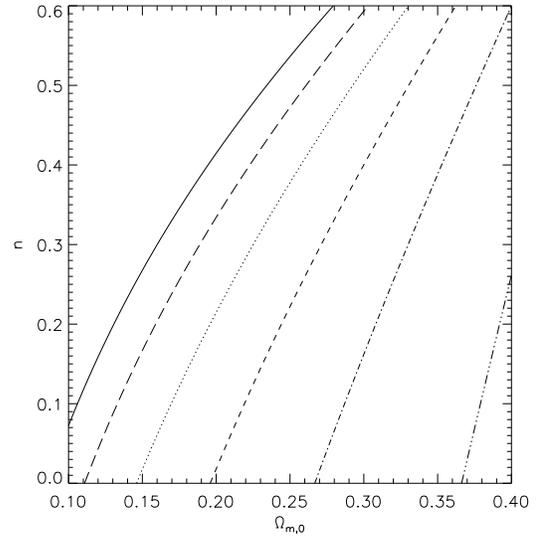}}
\caption{Contours of $z_{eq}$ with the Cardassian parameter, $n$, and $\Omega_m$. Contours for $z_{eq}$=1.2 (solid),
$z_{eq}$=1.0 (large-dashed), $z_{eq}$=0.8 (dotted), $z_{eq}$=0.6 (small-dashed), $z_{eq}$=0.4 (dot-dashed) and $z_{eq}$=0.2
(dot-dot-dashed) are shown.}
\label{fig:z_eq}
\end{center}
\end{figure}

\section{The Cardassian Model}

An additional term is added to the Friedmann equation such that

\begin{equation}
H^2=A\rho+B\rho^n
\label{equation:mfe}
\end{equation}

\noindent where $A$=$\frac{8\pi G}{3}$, $H$ is the Hubble constant, $\rho$ is the energy density and $n$ is a parameter of the  
Cardassian model. When the second term dominates, the scale factor $a$ $\propto~$t$^{\frac{2}{3n}}$, gives rise
to accelerated expansion for $n<\frac{2}{3}$. In order to determine $B$, it is convenient to equate the first  
and second terms at $z_{eq}$, the epoch at which the Cardassian term begins to dominate. Since
$\rho_m$=$\rho_{m,0}$(1+z)$^3$

\begin{equation}
B=\frac{8\pi G}{3}[\rho_{m,0}(1+z_{eq})^3]^{1-n}
\label{equation:b_exp}
\end{equation}

\noindent assuming the radiation contribution to the energy density is negligible. This expression gives rise to a 
modification in the critical energy density such that in a Cardassian cosmology

\begin{equation}
\rho_{crit}=F(n,z_{eq})\times\rho_{crit,old}
\label{equation:crit}
\end{equation}

\noindent where

\begin{equation}
F(n,z_{eq})=[1+(1+z_{eq})^{3(1-n)}]^{-1}
\label{equation:fact}
\end{equation}

\begin{figure}
\begin{center}
\centerline{\epsfxsize = 3in
\epsfbox{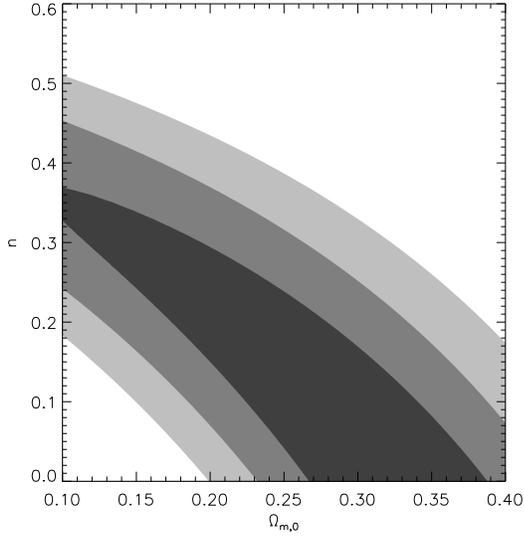}} 
\caption{Constraints from all 230 supernovae listed in Tonry et al. (2003) on $\Omega_m$ and the Cardassian parameter $n$.
The 1$\sigma$ (darkest shade), 2$\sigma$ and 3$\sigma$ confidence regions are shown.}
\label{fig:sne_all}
\end{center}
\end{figure}

\begin{figure}
\begin{center}
\centerline{\epsfxsize = 3in
\epsfbox{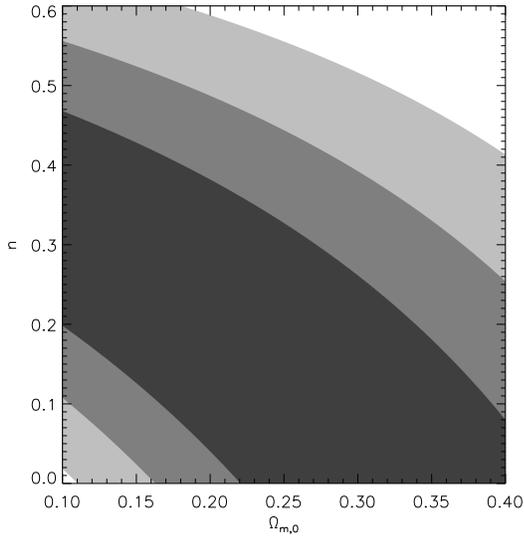}}
\caption{Constraints from the 54 supernovae published in Perlmutter et al. (1997) in Fit C, including supernovae from the
Calan-Tololo Survey and the Supernova Cosmology Project (taken from Tonry et al. (2003)), on $\Omega_m$ and the Cardassian
parameter $n$. The confidence regions are shown as before.}
\label{fig:sne_hzt}
\label{fig:sne_scp}  
\end{center}  
\end{figure}

\begin{figure}
\begin{center}
\centerline{\epsfxsize = 3in
\epsfbox{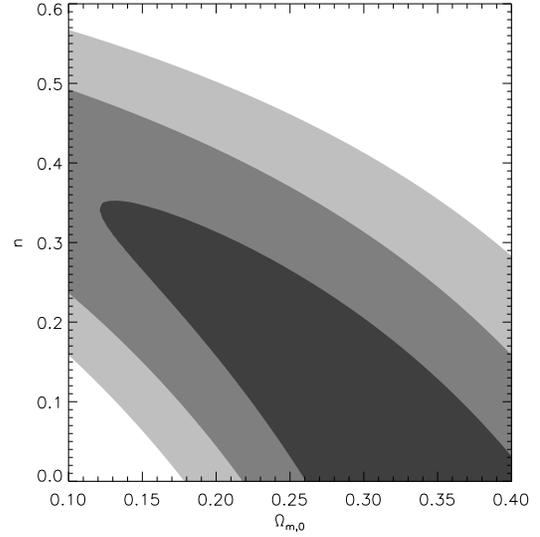}}
\caption{Constraints from the 130 supernovae published by the High-$z$ Supernova Search Team (taken from Tonry et al. (2003))
on $\Omega_m$ and the Cardassian parameter $n$. The confidence regions are shown as before.}
\label{fig:sne_hzt}
\end{center}
\end{figure}

\begin{figure}
\begin{center}
\centerline{\epsfxsize = 3in
\epsfbox{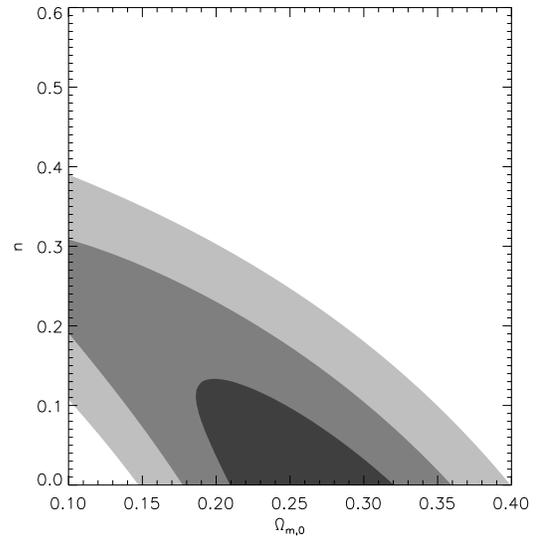}}
\caption{Constraints from the 172 supernovae (for which a redshift limit of $z>$0.01 and an extinction limit of
$A_V<$0.5 mag. were applied where possible) on $\Omega_m$ and the Cardassian parameter $n$. The confidence regions are shown
as before.}
\label{fig:sne_fit}
\end{center}
\end{figure}

\noindent which for a flat cosmology is equivalent to $\Omega_m$. Hence, this modification to the critical energy 
density is consistent with a flat, matter dominated cosmology. Fig.~\ref{fig:f_fact} shows the variation of $\Omega_m$
($\equiv$F(n,$z_{eq}$)) with $n$ and $z_{eq}$. Equivalently, Fig. ~\ref{fig:z_eq} shows the variation of $z_{eq}$ with 
$\Omega_m$ and $n$.

\section{Observational Constraints}

\subsection{Supernova Data}

In order to constrain the cosmological parameters of the Cardassian model, firstly, high redshift supernova data is used. 
Since altering the cosmology modifies the luminosity distance, the Cardassian model can be constrained by comparing 
empirical data with predicted values for the apparent magnitudes of standard candles at known redshifts. 
The relation between apparent and absolute magnitude can be written in the form (Perlmutter et al. 1997)

\begin{equation}  
m_B={\mathcal{M}}_B+5\log{\mathcal{D}}_L
\label{equation:appmag}
\end{equation}

\noindent where ${\mathcal{M}}_B$=$M_B-5logH_0+25$ is the Hubble constant-free absolute $B$-band magnitude, and is determined
empirically for type Ia supernovae as ${\mathcal{M}}_B$=-3.32$\pm$0.05 (Perlmutter et al. 1997) from the intercept on the 
Hubble diagram.${\mathcal{D}}_L$=$H_0d_L$ is the Hubble constant-free luminosity distance. The luminosity distance is given by

\begin{equation}
d_L(z)=\frac{c}{H_0}(1+z)\int_{0}^{z}\frac{dz}{E(z)}
\label{equation:lumdist}
\end{equation}

\noindent where $E(z)$ in a Cardassian cosmology is given by (Freese \& Lewis, 2002)

\begin{equation}
E^2(z)=\Omega_m(1+z)^3+(1-\Omega_m)(1+z)^{3n}
\label{equation:e_sq}
\end{equation}

Equations \ref{equation:appmag}, \ref{equation:lumdist} and \ref{equation:e_sq} determine a theoretical  
prediction for the apparent magnitude ($m_B(\Omega_m,n)$) of a supernova at a particular redshift.
Hence, $\Omega_m$ and $n$ are constrained through a determination of $\chi^2-\chi^2_{min}$ where 

\begin{equation}
\chi^2=\sum \frac{\left(m_B(\Omega_m,n)-m_{B_i}\right)^2}{\sigma^2_{m_{B_i}}}
\label{equation:chi_sq}
\end{equation}

\noindent where $\sigma_{m_{B_i}}$ is determined by quadrature of the errors on log${\mathcal{D}}_L$ and 
${\mathcal{M}}_B$. For a more robust determination of the errors, the uncertainty in ${\mathcal{M}}_B$ and the distance 
should be integrated over the entire parameter space, and an error on the redshift could also be included in the determination of 
$\chi^2$. However, these makes little difference to the resulting confidence regions and the determination of the errors 
described above is adequate for this analysis.

The observational data is taken from Tonry et al. (2003), which presents redshift and distance information
for 230 supernovae compiled from the literature and eight new supernovae from the High-$z$ Supernova
Search Team. Since the techniques for analysing the observational data vary between individual samples 
of supernovae, the authors have attempted to recompute the extinction estimates and the distance fitting using 
the methods of Riess et al. (1998), Jha et al. (2003) and Tonry et al. where possible, as well as apply a consistent 
zero-point for the local distance calibration. A complete description of this procedure is given in Tonry et al. (2003). 

Four supernova subsamples are selected from the 230 supernovae listed in Tonry et al. (2003). Firstly, the entire 
sample of 230 supernovae is used; the resulting 1$\sigma$, 2$\sigma$ and 3$\sigma$ constraints are shown in 
Fig.~\ref{fig:sne_all}. Secondly, the 54 supernovae included in Fit C in Perlmutter et al. (1997) taken from the 
Calan-Tololo Survey and the Supernova Cosmology Project are selected (see Fig.~\ref{fig:sne_scp}). Thirdly, the 130 
supernovae originally published by the High-$z$ Supernova Search Team are used to constrain the parameter space (see 
Fig.~\ref{fig:sne_hzt}). Finally,  a redshift cut of $z>$0.01 and an extinction cut of $A_V>$0.5 mag. is applied where 
possible, as in Tonry et al. (2003), in order to remove objects where an uncertain effect is apparent from either 
local peculiar velocities or host galaxy extinction; the constraints arising from the resulting 172 supernovae are shown in 
Fig.~\ref{fig:sne_fit}.
  
\subsection{The Cosmic Microwave Background}

In order to further constrain the available parameter space of the Cardassian model, empirical measurements for the 
locations of the first three Doppler peaks in the CMB angular power spectrum are compared with predicted values in a 
Cardassian cosmology. From Doran et al. (2001), the acoustic scale can be defined in terms of the conformal time

\begin{equation}
l_A=\pi \frac{\tau_0-\tau_{ls}}{c_s\tau_{ls}}
\label{equation:la_conf}
\end{equation}

\noindent where $\tau_0$ and $\tau_{ls}$ are the conformal time at the present day and at last scattering respectively.
c$_s$ defines the mean speed of sound before last scattering and is constant for a particular $\frac{\rho_b}{\rho_{rad}}$
and is taken to be 0.52 (Doran et al. 2001). The definition derived by Sen \& Sen (2003a) for this expression of l$_A$ in a
Cardassian cosmology is 

\begin{equation}
l_A=\frac{\pi}{c_s}\left(\frac{\int_{0}^{a_0}\frac{da}{X(a)}}{\int_{0}^{a_{ls}}\frac{da}{X(a)}}-1\right)
\label{equation:la_card}
\end{equation}

\noindent where $a_0$=1, $a_{ls}$=1100$^{-1}$ and

\begin{equation}
X(a)=\left(a+a^{4-3n}\left(\frac{1-\Omega_{m,0}}{\Omega_{m,0}}\right)\right)^{\frac{1}{2}}
\label{equation:x_fact}
\end{equation}

\noindent For a spectral index of n$_s$=1, the locations of the Doppler peaks are related to the acoustic scale by the relation

\begin{equation} 
l_m=l_A(m-\phi_m)
\label{equation:lm}
\end{equation}

\noindent where $\phi_m$ is a phase shift arising from driving and dissipative effects in the photon-baryon fluid
before last scattering, and is weakly dependent on $\Omega_m$, $\Omega_bh^2$ and n$_s$ (see Doran \& Lilley 2002). 
The expressions for $\phi_m$ for the first three acoustic peaks are given in Sen \& Sen (2003b). 
Hence for assumed values of $\Omega_bh^2$ and n$_s$, the locations of the Doppler peaks in a Cardassian 
cosmology can be predicted for a particular $\Omega_m$ and $n$. In the following analysis, values of 
$\Omega_bh^2$=0.02 (e.g. Burles et al. 2001; Pettini \& Bowen 2001; Kirkman et al. 2003) and n$_s$=1.0 are used, the effects of 
which are investigated in Sen \& Sen (2003b).

\begin{figure}
\begin{center}
\centerline{\epsfxsize = 3in
\epsfbox{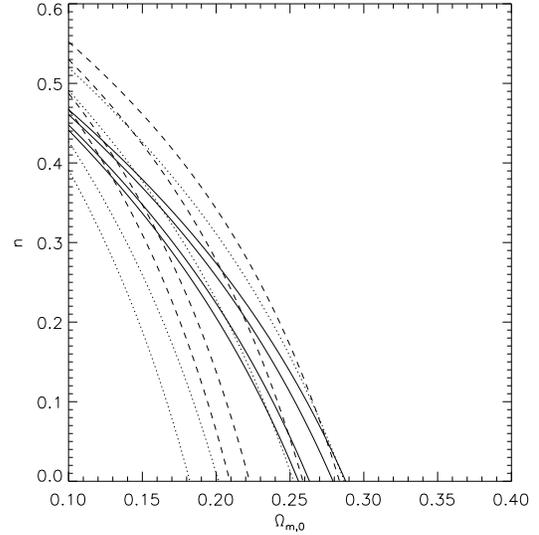}}
\caption{Constraints on $\Omega_m$ and the Cardassian parameter, $n$, from the location of the first (solid) and second 
(dotted) Doppler peaks in the WMAP angular power spectrum, and the third (dashed) Doppler peak from Boomerang. The 
1$\sigma$ and 2$\sigma$ contours are shown in each case.}
\label{fig:map}
\end{center}   
\end{figure}

The WMAP value for the locations of the first and second Doppler peaks are given as l$_1$=220.1$\pm$0.8 and
l$_2$=546$\pm$10 by Hinshaw et al. (2003) from Gaussian and hyperbolic fits to the power spectrum. The location 
of the third peak is given as l$_3$=825$^{+10}_{-13}$ (Hu et al. 2001). The resulting 1$\sigma$ and 2$\sigma$ 
constraints arising from each peak location are shown in Fig.~\ref{fig:map}.

\begin{figure}
\begin{center}
\centerline{\epsfxsize = 3in
\epsfbox{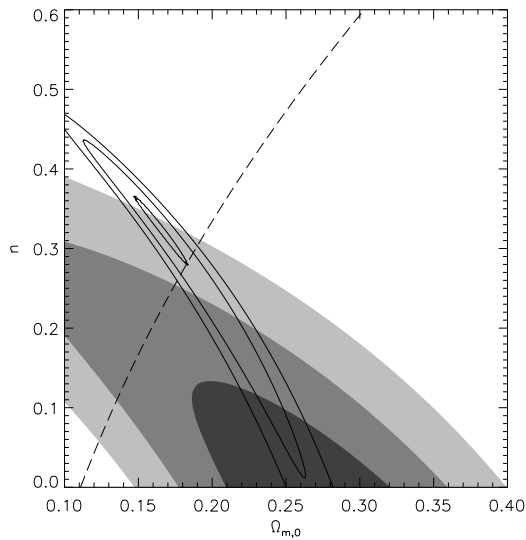}}
\caption{Here the constraints from the CMB data, the 172 supernovae sample (with $A_V<$0.5 mag. and $z>$0.01), 
and the upper limit on $z_{eq}$ are shown. The 1$\sigma$, 2$\sigma$ and 3$\sigma$ confidence regions from the 
first three Doppler peaks have been combined and are indicated by solid contours. The supernova confidence regions
are shown as in Fig.~\ref{fig:sne_fit}. The $z_{eq}$=1.0 track is indicated as a large-dashed contour as in Fig.~\ref{fig:z_eq}.}
\label{fig:all}
\end{center}
\end{figure}

\section{Discussion \& Conclusions}

The Cardassian expansion model utilises an additional term in the Friedmann equation, proportional to $\rho^n$, 
such that the Universe is flat, matter-dominated and in a phase of accelerated expansion. The need for a dark 
energy component is therefore removed, as the expansion observed in distant supernovae is driven solely by this 
additional component. In order to constrain the parameters of the Cardassian model $n$, $\Omega_m$ and 
$z_{eq}$ (the epoch at which the Cardassian term begins to dominate the expansion), it is necessary to select 
observational data free from any cosmological priors.

In this paper, various samples of type Ia supernova data from Tonry et al. (2003) were used to constrain 
$\Omega_m$ and $n$. Fig.~\ref{fig:sne_scp} shows the 1$\sigma$, 2$\sigma$ and 3$\sigma$ confidence levels 
derived from the 54 supernovae included in Fit C of Perlmutter et al. (1997). This is in general agreement with 
the constraints imposed by Sen \& Sen (2003a) for the same sample of supernovae, although the constraints are 
slightly broader, perhaps due to their use of contouring for absolute values of $\chi^2$. Since the errors are 
uncertain, more robust confidence levels can be derived when contouring around a minimum value of $\chi^2$ as used 
for the supernova data in this paper. Sen \& Sen (2003a) also use a different method of analysis in which 
a different value of ${\mathcal{M}}_B$ is used with an uncertain  associated error.

The tightest constraints from the supernova data are provided by the 172 supernovae sample (Fig.~\ref{fig:sne_fit})
for which an extinction limit of $A_V$=0.5 and a redshift cut of $z>$0.01 have been applied in order to remove objects with 
potentially large effects from perculiar velocities at low redshift and large host galaxy extinctions. The resulting 1$\sigma$ 
constraints favour a value of 0.19$<\Omega_m<$0.32 and $n<$0.13. However, the available parameter space within the 3$\sigma$ 
confidence 
region is large, although this represents a significant improvement over previous constraints. 

To constrain the available parameter space further, WMAP (Hinshaw et al. 2003) and Boomerang (Hu et al. 2001) 
measurements for the first three Doppler peaks in the CMB angular power spectrum are compared to the associated 
Cardassian prediction as in Sen \& Sen (2003b) (Fig.~\ref{fig:map}). These three constraints are in remarkably good 
agreement at $\Omega_m\sim$0.15, but begin to diverge at the values of $\Omega_m$ and $n$ suggested by the 172 supernovae sample.

In Fig.~\ref{fig:all}, the combined constraints from the first three Doppler peaks and the 172 supernovae sample are combined. 
Also included is the $z_{eq}$=1.0 track which represents an approximate upper limit arising from the effect of the Cardassian 
term in the modified Friedmann equation on local large-scale structure (Freese \& Lewis, 2002). The 1$\sigma$ constraint 
imposed by the combined CMB data indicates values of 0.15\lsim$\Omega_m$\lsim0.18, 0.28$<n<$0.36 and $z_{eq}>$1.0. However, 
there is good agreement at the 2$\sigma$ level with constraints of 0.19$<\Omega_m$\lsim0.26, 0.01$<n<$0.24 and 0.42$<z_{eq}<$0.89. 
This is consistent with the loose constraint of $z_{eq}<$1.0 imposed by local large-scale structure. However, it is in 
contradiction to the constraints derived by Sen \& Sen (2003) of 0.13\lsim$\Omega_m$\lsim0.23 and 0.31$<n<$0.44.

In conclusion, high redshift supernova data and recent mesaurements of the CMB power spectrum provide relatively tight constraints 
on the available parameter space of the Cardassian expansion model. There is a large discrepancy between the CMB and 
supernova 1$\sigma$ confidence regions, with the CMB 1$\sigma$ constraint implying a parameter space which is potentially at odds 
with local large-scale structure and is rejected by the supernova data at the $>$2$\sigma$ level. However, there is
good agreement at the 2$\sigma$ level, with the associated confidence regions limiting 0.19$<\Omega_m$\lsim0.26, 0.01$<n<$0.24 and 
0.42$<z_{eq}<$0.89.

\section*{Acknowledgments}
WJF would particularly like to thank Phil Outram, Mark Sullivan, Adam Myers, Tom Shanks and Ruth Emerson 
for useful discussion and technical assistance.

\label{lastpage}


\begin{thebibliography}{99}

\bibitem[\protect\citename{Burles et al. } 2001]{b1}
Burles, S., Nollett, K.M., \& Turner, M.S. 2001, ApJ, 552, L1

\bibitem[\protect\citename{Castillo-Morales \& Schindler } 2003]{b1}
Castillo-Morales, A. \& Schindler, S. 2003, A\&A, 403, 433

\bibitem[\protect\citename{Doran et al. } 2001]{b1}
Doran, M., Lilley, M., Schwindt,J. \& Wetterich, C. 2001, ApJ, 559, 501

\bibitem[\protect\citename{Freese \& Lewis } 2002]{b1}
Doran, M. \& Lilley, M. 2002, MNRAS, 330, 965

\bibitem[\protect\citename{Ettori et al. } 2003]{b1}
Ettori, S., Tozzi, P., \& Rosati, P. 2003, A\&A, 398, 879

\bibitem[\protect\citename{Freese \& Lewis } 2002]{b1}
Freese, K. \& Lewis, M. 2002, Phys.Lett.B, 540, 1

\bibitem[\protect\citename{Hinshaw et al. } 2003]{b1}
Hinshaw, G. et al. 2003, ApJ, submitted

\bibitem[\protect\citename{Hu et al. } 2001]{b1}
Hu, W., Fukugita, M., Zaldarriaga, M. \& Tegmark, M. 2001, ApJ, 549, 669

\bibitem[\protect\citename{Jha et al. } 2003]{b1}
Jha, S., Riess, A., \& Kirshner, R.P. 2003, in preparation

\bibitem[\protect\citename{Kirkman et al. } 2003]{b1}
Kirkman, D., Tytler, D., Suzuki, N., 0'Meara, J. \& Lubin, D. 2003, astro-ph/0302006

\bibitem[\protect\citename{Perlmutter et al. } 1997]{b1}
Perlmutter, et al. 1997, ApJ, 483, 565

\bibitem[\protect\citename{Pettini \& Bowen } 2001]{b1}
Pettini, M. \& Bowen, D.V. 2001, ApJ, 560, 41

\bibitem[\protect\citename{Riess et al. } 1998]{b1}
Riess, A.G. et al. 1998, AJ, 116, 1009

\bibitem[\protect\citename{Sen \& Sen } 2003a]{b1}
Sen, S. \& Sen, A.A. 2003a, ApJ, 588, 1S

\bibitem[\protect\citename{Sen \& Sen } 2003b]{b1}
Sen, S. \& Sen, A.A. 2003b, astro-ph/0303383

\bibitem[\protect\citename{Tonry et al. } 2003]{b1}
Tonry, J.L. et al. 2003, astro-ph/0305008

\bibitem[\protect\citename{Zhu \& Fujimoto } 2003]{b1}
Zhu, Z. \& Fujimoto, M. 2003, ApJ, 585, 52



\end{thebibliography}
\end{document}